# Room Temperature Operation of a Deep Etched Buried Heterostructure Photonic Crystal Quantum Cascade Laser

*R. Peretti[1]\*, V. Liverini[1], M.J. Süess[1], Y. Liang[1], P-B. Vigneron[1], J. M. Wolf[1], C. Bonzon[1], A. Bismuto[1], W. Metaferia[2], M. Balaji[2], S. Lourdudoss[2], E. Gini, M. Beck[1] and J. Faist[1]*

\*Corresponding Author: perettir@phys.ethz.ch

[1] Institute for Quantum Electronics, ETH Zürich, 8093 Zürich, Switzerland
[2] Laboratory of Semiconductor Materials, School of ICT, KTH-Royal Institute of Technology, 164 40 Kista, Sweden.

Abstract: High power single mode quantum cascade lasers with a narrow far field are important for several applications including surgery or military countermeasure. Existing technologies suffer from drawbacks such as operation temperature and scalability. In this paper we introduce a fabrication approach that potentially solves simultaneously these remaining limitations. We demonstrate and characterize deep etched, buried photonic crystal quantum cascade lasers emitting around a wavelength of 8.5 µm. The active region was dry etched before being regrown with semi-insulating Fe:InP. This fabrication strategy results in a refractive index contrast of 10% allowing good photonic mode control, and simultaneously provides good thermal extraction during operation. Single mode emission with narrow far field pattern and peak powers up to 0.88 W at 263 K were recorded from the facet of the photonic crystal laser, and lasing operation was maintained up to room temperature. The lasing modes emitted from square photonic crystal mesas with a side length of 550µm, were identified as slow Bloch photonic crystal modes by means of three-dimensional photonic simulations and measurements.

## 1. Introduction

Since their first demonstration in 1994, quantum cascades lasers (QCL) [1] have become a well-established semiconductor laser source emitting in a wide wavelength range from mid-infrared (MIR) to tera-hertz (THz). In the last decade the performance of QCLs was continuously improved by advances in active region design, [2, 3] layer processing [4, 5] and laser mounting. [2, 6] Meanwhile, the theoretical understanding of QCLs has progressed to a level where optimizations on very detailed levels can be performed. [7] Nowadays, QCLs are the most promising light source for many MIR laser-based applications e.g. gas sensing, [8, 9, 10, 11] process control [12] and biological sensing. [13] Applications such as military counter measure, [14] photo acoustics [15, 16] or surgery, [17] which specifically require high output power while keeping a very good far field, can be addressed by the DFB [18] or MOPA [19, 20] technology.

However, upscaling devices using these technologies is limited by the appearance of multimode behavior when increasing the ridge cross section beyond the laser wavelength or increasing the device length. A promising approach to circumvent these issues by keeping the coherence of a photonic mode on a large area is a two-dimensional (2D) periodic modulation of the refractive index at the scale of the wavelength, a so called photonic crystal (PhC). In the case of interband emitters, this strategy resulted in room-temperature (RT) emitters with a large surface area and output powers up to 3.5W, while keeping a high-quality, diffraction-limited output beam. [21] Consequently, the engineering of losses and modal overlap with the active region by means of PhCs, enables a fine control of the mode selection, while promising a significantly higher optical power output compared to conventional cavities.

Two approaches were investigated for QCLs so far. Initially, PhC QCLs with pillars separated by air holes [22] or polymer [23] were demonstrated at cryogenic temperature. These lasers profit from enormous refractive index contrast, but the use of plasmonic modes in the mid-infrared region induces large ohmic losses, preventing them to operate at room temperature. To relieve this issue, buried PhC DFBs based on standard buried heterostructure fabrication [4] were proposed as edge [24] and surface emitters [25] with large areas (100 µm x 3 mm) at a wavelength of 4.7 µm. However, despite these fabrication improvements, two major limitations remain:

1. The shallow-etched 2D PhC on top of the active region provides only a low index contrast, resulting in qualified as an "*extremely weakly coupled system*" [24] resulting in the emergence of parasitic modes, which cannot be suppressed by increasing the total size of the device.
2. While this fabrication strategy allows for high peak power (up to 12W [24]), RT operation and a diffraction limited far field, the operation of the device is limited to pulsed mode, which we accredit to an inefficient thermal management in the active region.

To address these two limitations, we implemented buried heterostructure (BH) PhC QCLs, where we deep dry etch the active region to form active region pillars, and then bury them by performing a regrowth of semi-insulating Fe doped InP (**Figure 1 A, B** and **C**). Firstly, this fabrication strategy provides strong photonic feedback due to a refractive index contrast of about 10%, which is of key importance to increase the coupling efficiency in order to maintain the spectral purity. Additionally, the strong index contrast ensures that the magnitude of index fluctuations resulting from process deviations or thermal instabilities during operation will be negligible compared to the index contrast itself, which is about $25 \times$ higher than the thermally induced index contrast during device operation ($\sim 4 \times 10^{-3}$). Secondly, the InP surrounding the active region is a good thermal conductor and acts as an exhaust pathway for the dissipated heat, which will eventually allow for higher operation temperatures and duty cycles. [4, 5]

**Figure 1:** A - Schematic of a BH PhC QCL mesa including buried photonic crystal of QCL active region (red), InP substrate, top cladding and lateral isolation (grey), as well as electrodes (gold). B - Scanning electron microscope image of a photonic crystal pillar exposed by focused ion beam (FIB) milling. C - 3D reconstruction of scanning electron microscope data recorded via FIB slice-and-view milling.

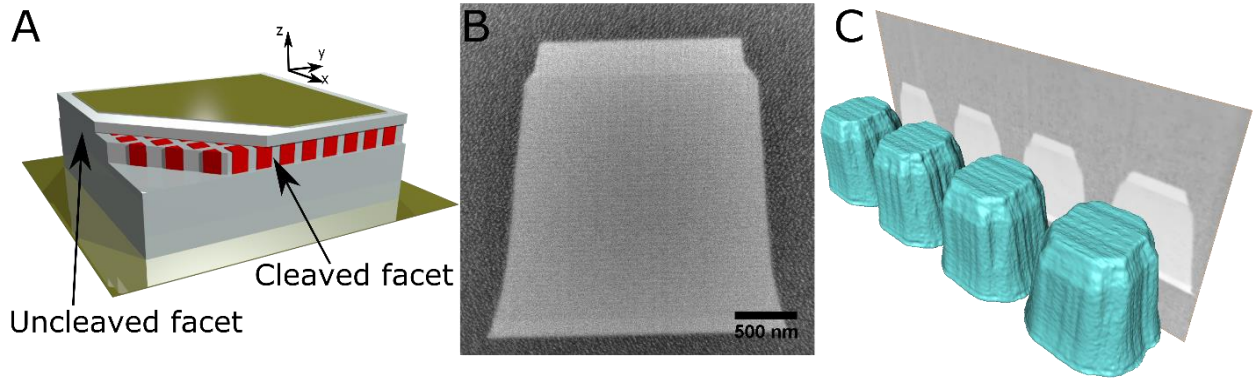

## 2. Experiments

Two active regions were grown, one based on lattice-matched (LM) as described in [26], the other one, adapted from [27], on strain-compensated (SC) InGaAs/AlInAs material system. Both are aimed at to operate at around 1200 cm$^{-1}$ (see **figure 2 B** and **C**). The SC active region previously showed better intrinsic performances [27], but is more difficult to process due to the strain in the individual layers. The active regions are sandwiched by 100 nm thick InGaAs layers. The processed PhC-QCL mesas are shaped as square arrays of micropillars, of period "*a*" around 2.7 µm and width about 2µm. The pillars were processed using deep-UV lithography followed by ICP dry etching. [28] The space between the etched pillars was filled with semi-insulating Fe-doped InP using a low pressure Hydride vapour phase epitaxy (HVPE). [28-30] A Si doped InP top cladding was regrown using Metal Organic Vapour Phase Epitaxy (MOVPE). [28] The top electrodes were formed using e-beam evaporated Ti/Pt/Au layers (10nm/40nm/250nm) in a lift off process. The back electrode is an e-beam evaporated Ge/Au/Ni/Au layer (see **figure 1 A**). Focused ion beam was used to obtain sliced cross sections that were then analyzed by scanning electron microscopy and are shown in **figure 1 B** and **C**. The structural parameters extracted from this analysis, are the period, *a = 2.7µm* and the areal filling factor of active regions ff=54% (corresponding to 74% in distance). From the growth parameters, the height of the active region and InGaAs claddings together is 2.5 and 2.55 µm for the LM and SC based structures, respectively. Samples containing square mesa PhC-QCLs of 200 periods with ~550 µm side length were mounted epi-side up on a copper sub-mount. The samples were either: -uncleaved or cleaved on -one, -two or -three facets along the x photonic crystal axis (see **figure 1 A**).

**Figure 2:** A - Computed band diagram, plotted along the Γ-X and Γ –M directions of the First Brillouin zone, plotted in reduced energy as a function of the in-plane wave vector. B - Electroluminescence spectrum taken with an FTIR spectrometer in scanning mode using an MCT detector and a lock-in amplifier. For the SC based device on the SC material cleaved on two orthogonal facets, the spectrum was recorded at 250K, 1% duty cycle, 12.2A for the luminescence and 14A for the laser. Spectra are plotted as a function of reduced energy units to indicate the energy spectrum where gain is expected. The photon energy (in cm$^{-1}$) is also indicated. The laser spectrum is indicated as a red line. C - Same data as B but for the LM device cleaved on one facet, here the laser spectrum is denoted by a blue line. The luminescence spectrum was recorded at 200 K, 1% duty cycle, 3.5 A for the luminescence and 4A for the laser. D - Experimental band diagram: normalized luminescence spectrum versus the in plane wave vector $k_y$ (One spectrum was taken every 10°(~0.01 π/a)). The computed band diagram (black lines) around the lasing point is superimposed (T = 15°C, period 2.8 µm, ff = 50%, mesa size 550µm, current 9.1 A). Inset: Schematic of the measurement description (cleaved facet denoted in black). E - Unfolded Brillouin zone of the photonic crystal. The yellow triangle depicts the direct coupling "light-cone" toward a facet which is cleaved perpendicular to the x direction (see **figure 1 A**).

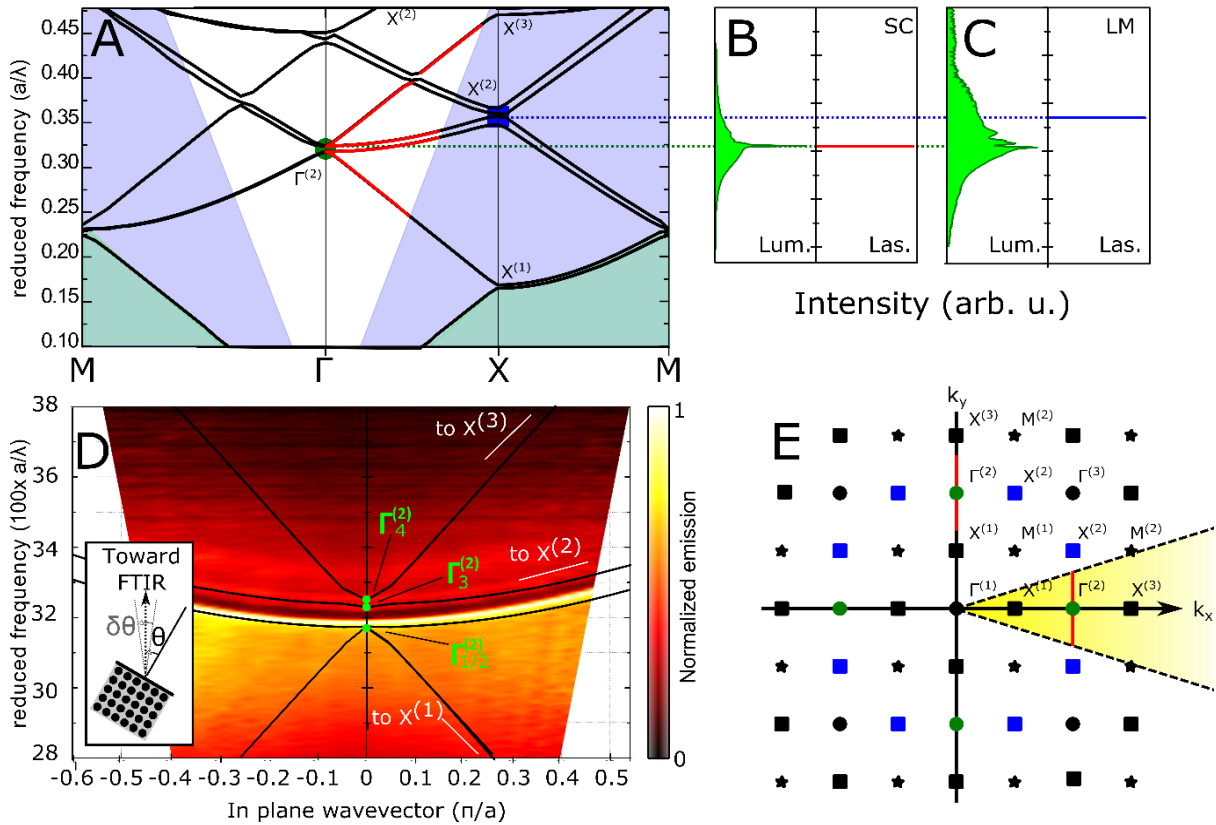

To analyze the modes supported by the PhC, three-dimensional (3D) finite difference time domain (FDTD) simulations were performed with the 3D FDTD Maxwell solver of the Lumerical © software package. [31] Bloch periodic conditions were used in the lateral direction while "perfectly matched layers" were used in the vertical direction. The material refractive index for InP (3.06) was taken from Palik [32]; for the active region it was calculated as an average of the refractive indices of the individual active region materials weighted by their total thickness (3.37). After the simulation, the spectral properties (frequency, quality factor) of the modes were extracted from the temporal signal using a harmonic inversion methods ("projection" of the temporal trace on the family of the damped sinus functions; Harminv [33]) as in [34]. This resulted in a band diagram plotted in **figure 2 A**. The $\Gamma^{(2)}$ k-point supports 3 modes all with sufficiently low horizontal losses to allow laser operation (cf. **table 1**) and $X^{(2)}$ k-point 4 other ones. **Figure 2 B** and **C** show the active regions can provide gain for both family of modes in $\Gamma^{(2)}$ and $X^{(2)}$ k-points. However $\Gamma^{(2)}$ modes have several advantages. First its frequency is at the maximum of the gain peak. Second in **figure 2 E** the facet emission cone defined by the Snell-Descartes law (the maximum angle is *a.sin(1/n$_{mode}$)*~18°) is depicted in yellow area delimited by black dashed lines. Thus, only this mode can be emitted through a facet in the classic square mesa geometry while X2 modes would require a specific geometry. Lasing action was actually observed only on $X^{(2)}$ for LM based device (see **figure 2 C** lasing on $X^{(2)}$ of a device cleaved on one facet) and on both k-points for the SC based devices (see **figure 2 B** lasing on $\Gamma^{(2)}$ of a device cleaved on two facets ). The details about modes in $\Gamma^{(2)}$ or $X^{(2)}$ selection are discussed in [28]). To better understand this laser operation, spontaneous emission and laser spectrum are shown on the same figure. The amplified spontaneous emission spectrum on **figure 2 C** contains a group of modes corresponding to modes originating from the crossing of bands at the $\Gamma^{(2)}$ k-point. Similar measurements were done on SC samples on **figure 2 B**. Again the amplified spontaneous emission spectrum contains a group of modes located around 0.32 $a^{-1}$ at the $\Gamma^{(2)}$ k-point. These modes are narrower than for the LM sample which is consistent with the fact that SC sample lases on $\Gamma^{(2)}$ modes while the LM lases on $X^{(2)}$ one.

To check if the parameters taken to calculate the frequency modes are accurate we measured the experimental band diagram shown on **figure 2 D**. Experimentally, the variation of $k_y$ vector corresponds to a rotation of the angle Θ around the growth axis. Sub-threshold luminescence spectra recorded at different Θ were normalized and plotted versus $k_y = \sin(\Theta) \, 2\pi / \lambda$. To better illustrate the match between angles and k vectors, the span of $k_y$ from **figure 2 D** is shown in **figure 2 A** and **2 E** by red lines. In **figure 2 A** three branches can be seen: one from $\Gamma^{(2)}$ to $X^{(1)}$, a double branch from $\Gamma^{(2)}$ to $X^{(2)}$ and the third from $\Gamma^{(2)}$ to $X^{(3)}$. These paths are shown in red in **figure 2 E** where only the double branch from $\Gamma^{(2)}$ to $X^{(2)}$ is within the facet emission cone and thus can directly couple through a facet. The two other branches require a lattice vector and thus couple less efficiently. The dispersion of the modes translates into a shift of the peak as a function of the observation angle. The resulting curves exhibit parabolic dispersion behavior (non-linear dispersion term and curvature of 0.036 $a^{-1}/(\pi/a)^2$), a usual signature for a slow Bloch mode. For comparison, the band diagram of **figure 2 A** is superimposed. The good match between the experiments and the simulations shows the validity of the input parameters of the simulation. The superposition shows as well that the two modes forming the gap are the modes $\Gamma^{(2)}_{1/2}$ and $\Gamma^{(2)}_{3}$. Since the lasing action occurs on the mode with the lowest energy (see **figure 2 C**), the lasing mode is unambiguously identified as $\Gamma^{(2)}_{1/2}$. In addition, the bands from $\Gamma^{(2)}$ to $X^{(2)}$ are visible in the experimental data, but not those from $\Gamma^{(2)}$ to $X^{(1)}$ or $X^{(3)}$, which is explained by the fact that the band to $X^{(2)}$ can couple directly through a facet cleaved perpendicular to x (see **figure 1 A**), but the one to $X^{(1)}$ or $X^{(3)}$ only by adding a grating vector (cf. **figure 2 E**).

To better understand why the mode in $\Gamma^{(2)}_{1/2}$ lases predominantly, 3D-FDTD simulations were conducted using the structural information extracted from the FIB measurement displayed in **figure 2 C**. In order to quantify the modal losses and overlap factors, we take into account the vertical confinement by calculating the field distribution of each mode. In addition, the horizontal losses were evaluated by computing 2D simulations of finite size sample (200 periods) in both x and y directions. For these simulations the PhC was surrounded by "perfect match layers" (absorbing boundary conditions) on each side, for the case of only one or less cleaved facet.

**Figure 3:** A - bar diagram of losses expressed in threshold gain (losses normalized by the overlap) for the three modes in $\Gamma$. B - Cross-section of the z-component of the electric field distribution of each of these modes

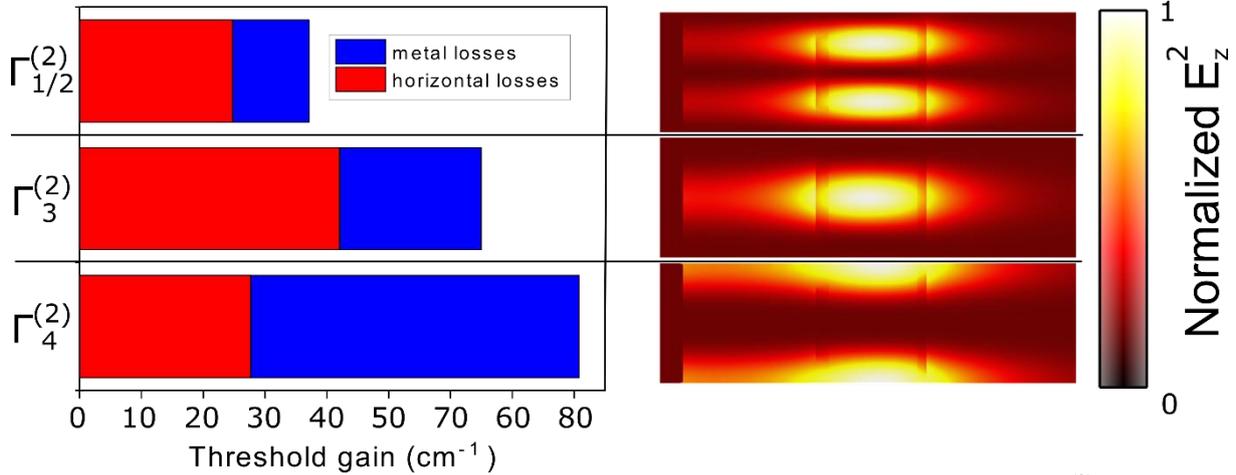

For infinite structures without a metal layer the simulations predict modes at the $\Gamma^{(2)}$ k-point with vertical losses negligible compared to free carrier absorption (FCA) (~4 cm$^{-1}$) and thus to the actual losses in the device. Hence, vertical losses would not play any role in the mode selection in such structures. However, as shown in **figure 3 A & B,** in finite structures the computed horizontal losses and metal losses are higher than losses from FCA and will play a role in the selection process. In fact the $\Gamma^{(2)}_{1/2}$ mode has the highest overlap factor among all the modes in $\Gamma^{(2)}$ (0.29 compare to 0.21 for mode $\Gamma^{(2)}_3$ and 0.13 for mode $\Gamma^{(2)}_4$), which has a twofold effect. First, the higher index contrast improves the waveguide confinement as shown in **figure 3 B**. This directly impacts the metal losses as shown in **figure 3 A**. Second, the gain is higher than the one imparted to other modes at the same voltage. All in all, since the actual losses are the sum of the PhC losses and the absorption losses (both the FCA and those from the top electrode), laser operation occurs at mode $\Gamma^{(2)}_{1/2}$ because the maximum gain of the active region (~ 70 cm$^{-1}$, estimation from data in [27]) cannot overcome the losses of the other modes.

**Figure 4:** A - LIV curve for a PhC laser and FP laser of the same structure (SC based lasers): the current-voltage curve uses the left ordinate and the light-current curve the right ordinate). PhC laser characteristics: period=2.6 µm, ff=54%, 530 µm mesa; FP characteristics: 2 mm long and 12 µm wide. Duty cycle 0.025% and repetition rate 2.5 kHz. Inset spectrum of the PhC laser at 263 K and 14A. B - Far field patterns of the lasing modes $\Gamma_{1/2}^{(2)}$ of PhC. The horizontal angle rotation is the rotation around axis z (the growth direction).

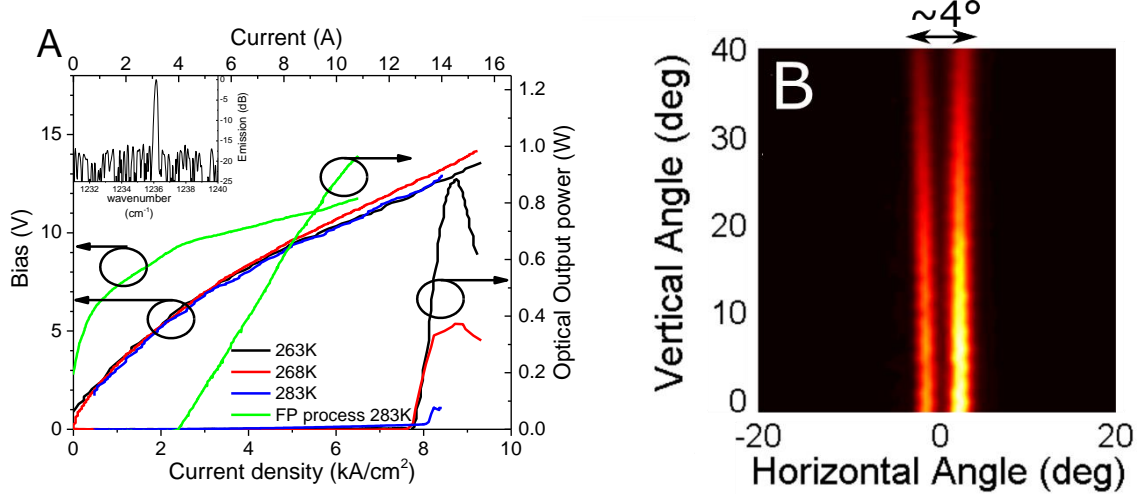

The power-current-voltage (LIV) characteristics of an SC laser were recorded at a repetition rate of 2.5 kHz and a duty cycle of 0.025% (**figure 4 A**). The power was measured with a liquid nitrogen cooled MCT detector (using a density filter while needed) and calibrated with a power meter. The characteristics of an uncoated 3 mm long 12µm wide Fabry-Perot (FP) device from the same MBE growth (recorded at 288K) with similar process as [5, 26] are plotted in the same figure, in order to compare to the threshold losses of the BH PhC devices. The Bias and current density scale are the same but current and power axis in are not to scale for the FP device. The facet emitting mode $\Gamma_{1/2}^{(2)}$ reaches peak powers of up to 0.88 W. The voltage at threshold is about 1V higher in the case of PhC process. It is explain by the fact that the loss on the FP process is more than 2 time lower (~ 15cm$^{-1}$) than in the PhC process. In addition, the IV curves show some leakage in the PhC process compared to the FP laser of the same structure. This effect is most pronounced in the SC based devices, where typical threshold currents around 7.8 kA/cm$^2$ are observed, in contrast to 3 kA/cm$^2$ for the LM based devices (instead of 2.7 kA/cm$^2$ around 270 K for SC based device and 2.5 kA/cm$^2$ at 250 K for LM based device, c.f. [27]). This difference can be explained by the fact that the growth rate at the beginning of the lateral regrowth of InP:Fe was higher than targeted for the SC process. This can possibly lead to an incomplete incorporation of iron.

A far field pattern recorded on a PhC sample from the SC process is shown in **figure 4 B**. The horizontal and vertical angles correspond to rotations around the z- and x–axis displayed in **figure 1 A**, respectively. The mode at $\Gamma^{(2)}_{1/2}$ shows two narrow lobes. The full width at half maximum corresponds to the resolution limit of the setup coinciding with the diffraction limited low spatial-frequencies pass filter of a 550 µm long facet (~1.8°). The two lobes are separated by a divergence angle of about 4°. The $\Gamma^{(2)}_{1/2}$ mode coupling through a facet has a k-vector perpendicular to it (see **figure 2 E**). This means that the phase of the wave on this facet is not modulated by the Bloch mode of the PhC. Thus the only modulation giving rise to double lobe behaviour is a modulation by the envelope of the PhC Bloch mode. The angle between the two lobes corresponds to the diffraction limited emission of the second order lateral mode (i.e. antisymmetric mode with respect to the centre of the emission facet). The spectral difference between this mode and the fundamental mode is ~ 0.01cm$^{-1}$ which cannot be resolve with our apparatus, however the far field pattern clearly show single mode behaviour (c.f. [28]). The broad angle in the vertical direction corresponds to the diffraction limited angle through a 3 µm thick slit (thickness of the active region).

## 3. Conclusion

In summary, we presented a technological approach implementing deep-etched photonic crystals in QCLs. This strategy will eventually become capable to overcome the contemporary limitations to PhC QCL device performances with respect to thermal management and mode control. By deep etching and burying the QCL active region, we demonstrated PhC QCLs lasing single mode at 2 different symmetry points of the PhC Brillouin zone. The back-filling InP:Fe regrowth between the active region pillars employed during the fabrication will enable better heat extraction in such devices, while simultaneously providing a large refractive index contrast for improved mode control. This is evidenced by room temperature single-mode operation for both investigated processes. The highest performance was achieved for a > 550 µm x 550 µm QCL lasing single-mode with 0.88 W output power per facet.

While the conventional DFB PhC QCL process [24, 25] has reached maturity, the devices resulting from our buried PhC QCL process, while still in its infancy, can already compete with state of the art performances, especially in maintaining spectral purity and good far field pattern at all operation currents. Finally, it harbours the prospect of surpassing previous results also in terms of thermal management, once the technology has fully matured, which will pave the way for large-area high-power single-mode QCLs capable of fulfilling the demanding specifications posed by high end applications.

**Acknowledgements** This work was supported by EU FP7 MIRIFISENS Integrated Project 317884 and the Swedish Research Council (VR) through the grant proposal 2015-05470 and Linné Center of Excellence, ADOPT.

Received: ((will be filled in by the editorial staff))
Revised: ((will be filled in by the editorial staff))
Published online: ((will be filled in by the editorial staff))

**Keywords**: quantum cascade laser, photonic crystal, mid infrared, single mode

**Graphical Abstract**

The table of contents (ToC) entry should be up to 70 words long. The entry should be written in the present tense and impersonal style. The text should be different from the abstract text.

This paper presents the demonstration and characterization of deep etched square lattice photonic crystal quantum cascade laser emitting at around 8.5 µm under room temperature operation. The observed lasing modes were identified as slow Bloch photonic modes of a 550 µm side square photonic crystal mesa. The measured emission showed an output peak power of 0.88 W at 263 K with single mode behavior and narrow far field pattern, from a facet.

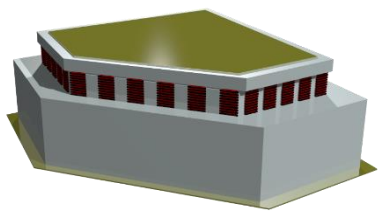

ToC figure ((Please choose one from your article or one specifically designed. Size: about 50 mm broad × 50 mm high. No caption.))